\documentclass[prx,twocolumn,date,floatfix,preprintnumbers,amsmath,amssymb,superscriptaddress]{revtex4}

\usepackage{graphicx}
\usepackage{amsmath}
\usepackage{amssymb}
\usepackage{color}
\usepackage{dcolumn}
\usepackage{epsfig}
\usepackage{bm}
\usepackage{subfigure}
\usepackage[urlcolor=blue]{hyperref}
\hypersetup{backref, colorlinks=true, linkcolor=blue, citecolor=blue}

\begin{document}

\title{Full Set of Superconducting Parameters of K$_3$C$_{60}$}

\author{Ren-Shu Wang}
\affiliation{School of Science, Harbin Institute of Technology, Shenzhen 518055, China} 
\affiliation{Center for High Pressure Science and Technology Advanced Research, Shanghai 201203, China}

\author{Di Peng}
\affiliation{Center for High Pressure Science and Technology Advanced Research, Shanghai 201203, China}

\author{Li-Na Zong}
\affiliation{Center for High Pressure Science and Technology Advanced Research, Shanghai 201203, China}

\author{Zeng-Wei Zhu}
\affiliation{Wuhan National High Magnetic Field Center, Wuhan 430074, China}

\author{Xiao-Jia Chen}
\email{xjchen2@gmail.com}
\affiliation{Center for High Pressure Science and Technology Advanced Research, Shanghai 201203, China}
\affiliation{School of Science, Harbin Institute of Technology, Shenzhen 518055, China} 

\date{\today}

\begin{abstract} 
The superconducting parameters are the key for building or identifying the theory responsible for the mechanism of superconductivity. Such parameters for fulleride superconductors have not been well established despite the tremendous efforts over the past 30 years. Here we provide a full set of parameters through a systematic study on a well-characterized K$_{3}$C$_{60}$ sample. The obtained high upper critical field of 33.0$\pm$0.5 T from the direct electrical transport measurements together with the relatively high critical temperature and large critical current density classifies K$_{3}$C$_{60}$ as a promising three-dimensional superconducting magnet material with the advantage of the rich carbon abundance on the Earth. This high upper critical field along with the large reduced superconducting energy gap and strong phonon self-energy effect supports the strong electron-phonon coupling interactions in this superconductor. The evaluation of all self-consistently obtained parameters suggests the unconventional nature of the superconductivity for K$_3$C$_{60}$ with the joint contributions from the strong electron-phonon coupling and electron correlations. These results and findings are important not only for fundamentally understanding the superconductivity in fullerides but also for future superconducting magnet developments and applications.
\end{abstract}

\maketitle

\section{Introduction}
The discovery of superconductivity in alkali fullerides with the critical temperature $T_c$ going from 18-19 K for K$_3$C$_{60}$ \cite{Heb} through 28-29 K for Rb$_3$C$_{60}$ \cite{ross} to 38-40 K for Cs$_3$C$_{60}$ \cite{pals,taka,Gan} is an important event in modern science after the birth of the new form of carbon called `buckminsterfullerene' or in short fullerene \cite{krot}. Apparently, these three-dimensional molecular solids are different in structure from the early discovered high-$T_c$ cuprates and recently discovered twisted graphene \cite{ycao}. They in reality share the similar superconducting phase diagram emerging from neighbouring Mott insulating state \cite{taka,Gan,ycao}. This similarity fuels the hope to solve the long-standing puzzle of the mechanism of superconductivity in cuprates once knowing the key factors that govern superconductivity in fullerides due to the relatively easy realization of superconductivity with high $T_c$ in fullerides compared to the need of precise tuning in graphene, another purely C-based superconductor.  

The crucial examination of the theory highly depends on the superconducting parameters. Take the first fullerene superconductor K$_{3}$C$_{60}$ as an example, the large differences of these parameters from the experiments make the comparisons with the theory difficult \cite{Gun0,capo}. Conducting electrical transport measurements to realize the zero resistance state, known as one of the two essential features of a superconductor, has long been a challenge in the characterization of its superconductivity. Therefore, the characterization of fullerene superconductors has mainly been performed based on magnetization measurements \cite{Heb,Ste,Hol,Yoo} due to the challenges in handling the air-sensitive samples \cite{Fle,Hol1}. Among the several electrical transport studies on K$_3$C$_{60}$ \cite{Pal,Xia,Kle}, magnetic characterizations were absent. In fact, the first experimental evidence for supporting superconductivity in K$_3$C$_{60}$ from the zero-resistance state and Meissner effect was obtained from different samples \cite{Heb}. The airtight devices were developed to measure the resistivity for films \cite{Heb,Pal,Pal1,Had1} and single crystals \cite{Zha,Mar} with the observed zero-resistance state in the superconducting state.  Meanwhile, some residual resistances often appeared at low temperatures in the electrical transport measurements on single crystals \cite{Xia,Xia1,Hou}, probably due to the limited crystallinity in films \cite{Heb,Pal,Pal1} or uncontrolled alkali distribution in single crystals \cite{Xia,Xia1,Hou}. Alternatively, the resistivity was inferred from some contactless methods \cite{Deg,Kle}. However, these techniques brought about large residual surface resistance \cite{Kle}. In view of these facts, the electrical transport measurements are still challenging in the characterization of superconducting fullerides. The zero-resistance state and Meissner effect on the same sample have only been realized recently for K- and Rb-doped C$_{60}$ through the improvements of the sample quality and measurement techniques \cite{rswang,zong}. 

It has been generally accepted that the resistivity measurement at required high magnetic field and low temperature is a reliable method for accurately determining the critical magnetic field which is fundamentally important due to its close relation to the coherent length and the Cooper pairing strength \cite{Tin}. At the moment, the upper critical field $H_{c2}$(0) values of K$_3$C$_{60}$ were obtained either through the low-magnetic-field measurements of the magnetization \cite{Hol,Bun} and resistivity \cite{Hou,Hou1,Pal} or the high-magnetic-field non-contact measurements based on the $ac$ magnetization \cite{Joh,Boe,Bae} and radiofrequency technique \cite{Fon,Kas}. Having accurate $H_{c2}$(0) of K$_3$C$_{60}$ through the resistivity measurements at high enough magnetic fields is still highly desired. Here we choose K-doped C$_{60}$ with the considerations of its similarity to other fullerides \cite{Fle,Gun0} and simplicity due to the absence of the antiferromagnetic state to solve this long-standing issue with the purpose to provide the constraints on the identification of the existing theoretical models or the future theory developments from a full set of superconducting parameters. 

\begin{figure}[tbp]
\includegraphics[width=1.0\columnwidth]{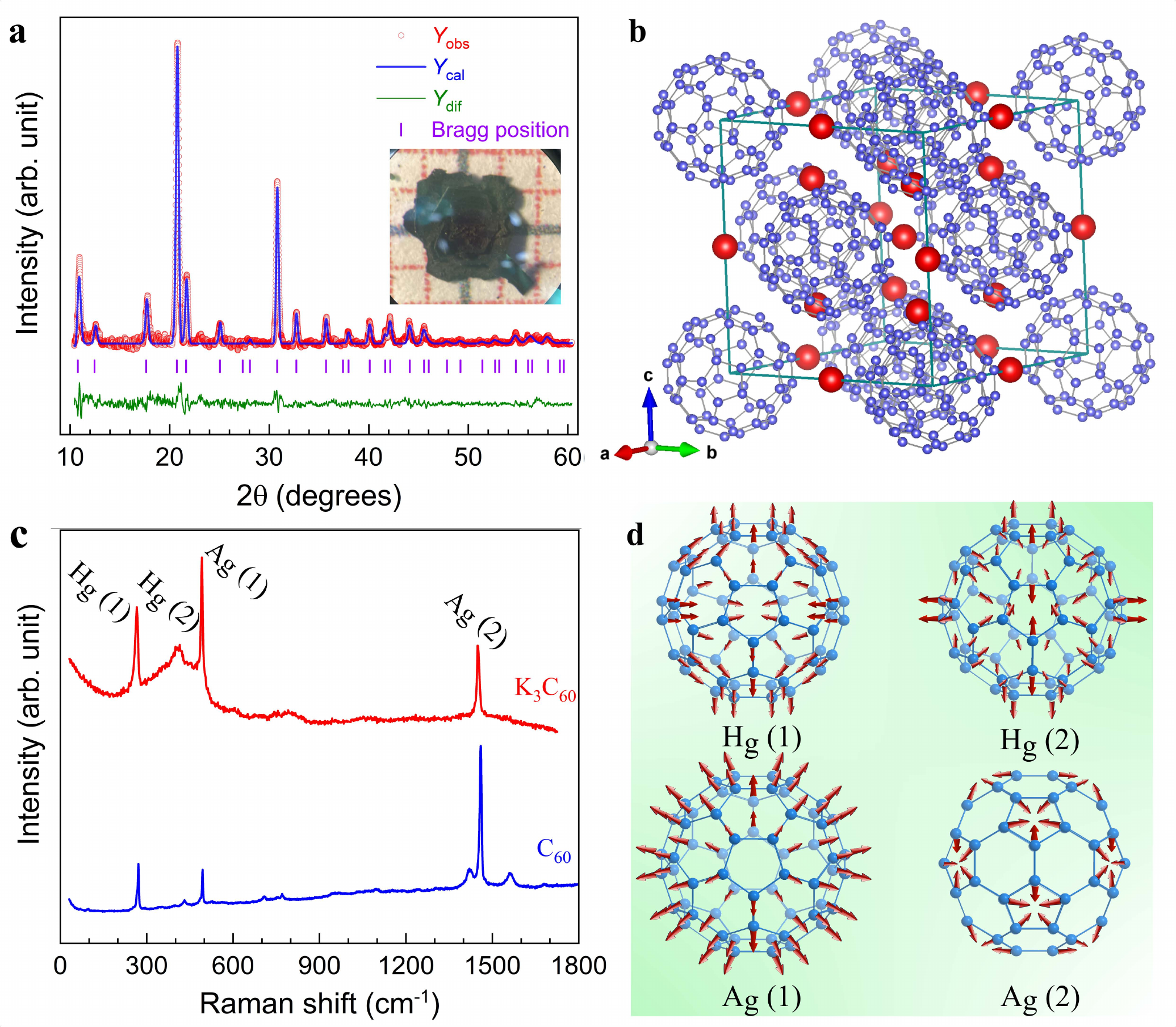}
\caption{Structural and spectroscopic characterizations of K-doped C$_{60}$. (a) Experimentally observed (red dots) and calculated (blue solid line) x-ray diffraction patterns for the doped sample at room temperature. The vertical purple lines present the Bragg reflection positions of the $fcc$ structure, and the solid green line at the bottom represents the difference profile. Inset: Sample picture taken under microscope. (b) Crystal structure of $fcc$ K$_3$C$_{60}$. The violet and red balls represent C and K atoms, respectively. (c) Raman scattering spectra of pure and K-doped C$_{60}$. (d) Schematic diagram of four representative intramolecular vibrational modes for K$_3$C$_{60}$. The red arrows indicate the vibrational directions of the C atoms.}
\end{figure}

\section{Results}

\subsection{Crystal structure and molecular vibrations of K-doped C$_{60}$}
  
Wet chemistry method was used to synthesize K-doped C$_{60}$ detailed in Method. The quality, structure, and phase of the sample were examined by x-ray diffraction (XRD) and Raman scattering measurements. The XRD profile can be well indexed by the $Fm\overline{3}m$ space group [Fig. 1(a)], indicating that the synthesized K-doped C$_{60}$ sample is a single-phase compound with the face-centered cubic ($fcc$) structure [Fig. 1(b)]. The obtained lattice parameter $a$ is 14.22$\pm$0.01 {\AA}, in good agreement with those reported previously \cite{Ste,Fle}. 

Figure 1(c) shows Raman spectra of the pristine and K-doped C$_{60}$ collected at room temperature. All the Raman-active modes for C$_{60}$, including two $A_g$ modes (at 497 and 1469 cm$^{-1}$) and eight $H_g$ modes: 272, 432, 710, 772, 1100, 1248, 1422, and 1574 cm$^{-1}$, are observed in the pristine C$_{60}$, agreeing well with the literature \cite{Zhou}. The two low-frequency $H_g$ modes and two $A_g$ modes have strong intensities in K-doped C$_{60}$. The vibrations of these four modes \cite{schlu} are shown schematically in Fig. 1(d). A recognized approach to determine the doping level of K$_x$C$_{60}$ is the line-shift of the pinch mode $A_g$(2). The observed 17 cm$^{-1}$ redshift of this mode yields the doping concentration of about 3 estimated by the empirical relation of 6 cm$^{-1}$ per elementary charge \cite{Had}. Therefore, the synthesized sample is homogeneous and has the chemical formula of K$_3$C$_{60}$ with the $fcc$ structure.

\begin{figure}[tbp]
\includegraphics[width=\columnwidth]{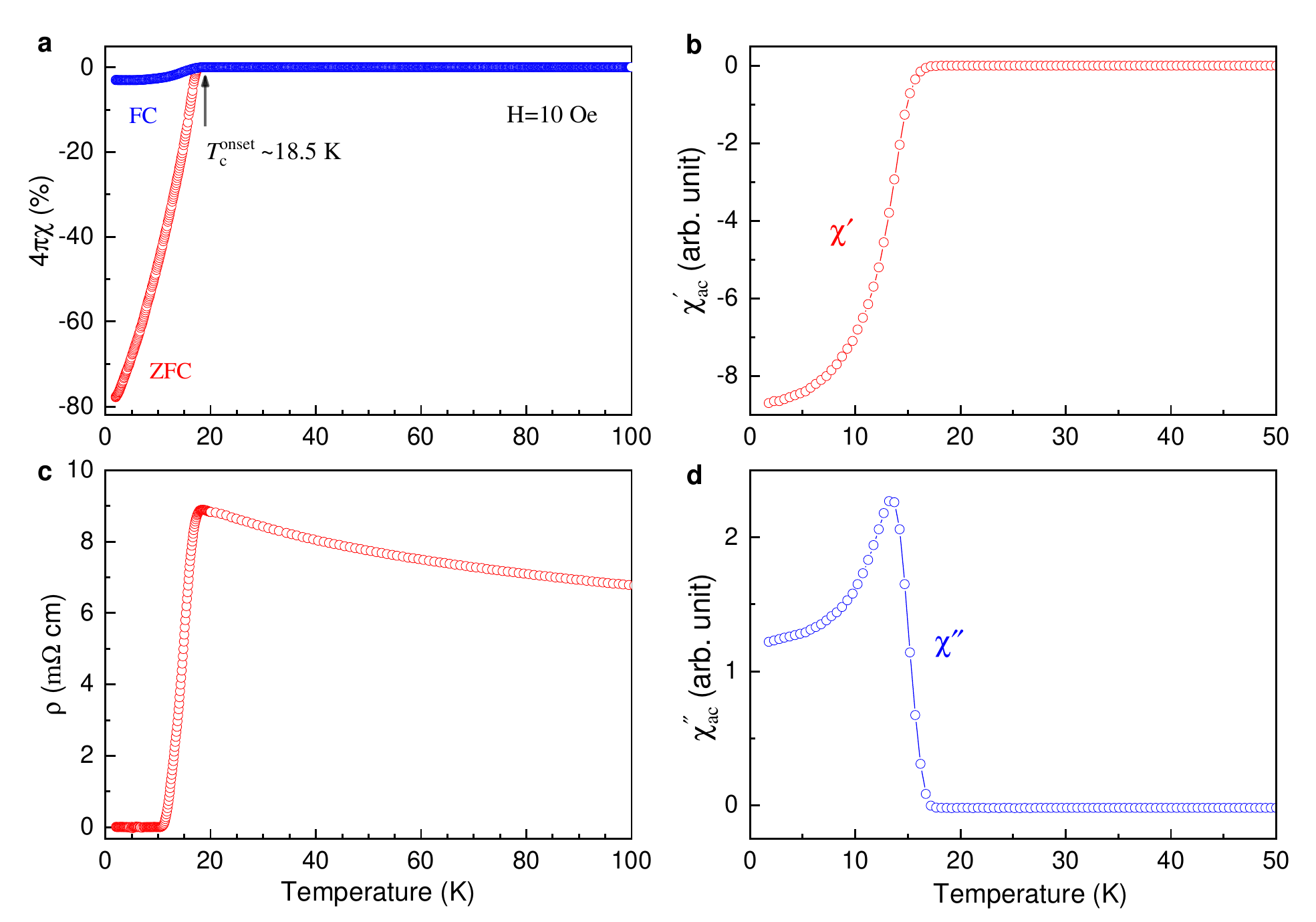}
\caption{Characterization of superconductivity of K$_{3}$C$_{60}$ from the magnetic susceptibility and electrical transport measurements. (a) $\emph{DC}$ magnetic susceptibility ($\chi$) with temperature range of 1.8-100 K at the applied magnetic field of 10 Oe. Zero-field-cooling (ZFC, red circles) and field-cooling (FC, blue circles) $\chi$ show an obvious diamagnetism below 19 K. (b) Real part of the $\emph{ac}$ magnetic susceptibility ($\chi^{\prime}_{ac}$) in the temperature range from 1.8 to 50 K. The probe harmonic magnetic field and frequency are 5 Oe and 234 Hz, respectively. (c) Temperature dependence of the electrical resistivity ($\rho$). (d) Imaginary component of the $\emph{ac}$ magnetic susceptibility ($\chi^{\prime\prime}_{ac}$) as a function of temperature (1.8-50 K).}
\end{figure}

\subsection{Superconductivity of K$_{3}$C$_{60}$}
 
The superconductivity of the synthesized K$_3$C$_{60}$ was identified by the existence of both the Meissner effect and zero resistance state. Figure 2 shows the results from the $dc$ and $ac$ susceptibility ($\chi$) and resistivity measurements. The $dc$ susceptibility directly demonstrates the Meissner effect of the sample, as shown in Fig. 2(a). Zero-field-cooling (ZFC) and field-cooling (FC) $\chi$ curves exhibit a clear drop below 18.5 K, which is defined as $T_c$. The shielding fraction (SF) of 80\% at 2 K is estimated from the ZFC $\chi$ curve. This diamagnetic effect is also observed through the real part ($\chi^{\prime}_{ac}$) of the $ac$ susceptibility [Fig. 2(b)], similar to the ZFC $\chi$ curve from the $dc$ measurement. As shown in Fig. 2(c), the resistivity drops rapidly below 18.7 K and quickly reaches the absolute zero value. The zero-resistivity transition of the sample is also indirectly reflected by the imaginary part ($\chi^{\prime\prime}_{ac}$) of the $ac$ susceptibility. To response the vortex current signal, $\chi^{\prime\prime}_{ac}$ can only be induced when the resistivity drops sharply because of the relatively large resistivity in the normal state and the flux exclusion in the complete Meissner state \cite{Gom}. The peak shape of $\chi^{\prime\prime}_{ac}$ in Fig. 2(d) is the manifestation of the sharp drop in resistivity. As learned from the literature for the representative magnetization \cite{Heb,Ste,Hol,Hol1,Yoo,Boe,Bae} and nearly all electrical transport \cite{Heb,Pal,Uga,Zha,Hes,Mar,Xia,Xia1,Hou,Pal1} measurements, the evidence for the zero-resistance state and the Meissner effect taken on the same K$_3$C$_{60}$ sample is quite rare. The present study together with our recent efforts \cite{rswang,zong,rswang2} provides solid experimental evidence for supporting superconductivity in synthesized fullerides based on its two essential characters.

\subsection{Upper critical field determined by resistivity measurements}

Figure 3(a) shows the temperature dependence of the resistivity at low magnetic fields of 0-9 T. With increasing magnetic field, $T_c$ shifts toward lower temperatures [Fig. 3(a)]. The resistivity behaviours at low temperatures and pulsed magnetic fields up to 50 T are given in Fig. 3(b). The determination of $H_{c2}$ at a given temperature is illustrated in the inset of this figure. The $H_{c2}(T)$ data obtained from the high- and low-magnetic fields can be well described by the Werthamer-Helfand-Hohenberg theory \cite{Wer}. Note that the experimentally observed $H_{c2}$'s at low temperatures are significantly larger than the orbital limit field $H_{c2}^{orb}$(0) of 22 T estimated from the $H_{c2}(T)$ slope in the low-field regime ($H \le$1 T) by using $H_{c2}^{orb}$(0)=0.69$\times$$T_c$$\times\lvert$d$H_{c2}$/d$T$$\rvert$$_{T=T_c}$, suggesting the strong coupling effect. The experimentally obtained $H_{c2}$(0) of 33.0$\pm$0.5 T basically agrees with most of the high-field experimental results for this material \cite{Joh,Boe,Fon,Kas}.

The reliability of the obtained $H_{c2}$(0) value is guaranteed by the direct electrical transport measurements at enough high magnetic fields and at low temperatures. As a key indicator in the applications of superconducting materials, the $H_{c2}$(0) of K$_3$C$_{60}$ is little bit higher than that of Nb$_3$Sn with lower $T_c$ of 18 K \cite{God}, a typical three-dimensional superconductor well known as superconducting magnets. Interestingly, our $H_{c2}$(0) value for K$_3$C$_{60}$ at ambient pressure is even comparable to those of $fcc$ CeH$_{10}$ with much higher $T_c$=115 K at pressure of 95 GPa \cite{Che} and ThH$_{9}$ with $T_c$=151 K at 170 GPa \cite{seme}. However, the need of high pressures to access the superconducting state in these superhydrides limits their technological applications. The larger $H_{c2}$(0) for K$_3$C$_{60}$ with the higher $T_c$ compared to those in Nb$_3$Sn offers the attractive opportunities for the technological applications due to the rich carbon abundance on the Earth as well as the low cost. Therefore, the present experimental finding opens a window for future developments and designs of the three-dimensional superconducting magnets based on this material. 

\subsection{Lower critical field, penetration depth, and coherence length}

The obtained $H_{c2}$(0), combined with other superconducting parameters such as the lower critical field $H_{c1}$, London penetration depth $\lambda_L$, and coherence length $\xi$, are crucial to the understanding of the physical properties and the mechanism of superconductivity of a type-II superconductor. The $H_{c1}$ is determined by the magnetic-field dependence of the $dc$ magnetization $M(H)$ at various temperatures below $T_c$ [Fig. 3(c)]. Within this method, the magnetic field that initially deviates from the linear $M(H)$ behaviour is defined as $H_{c1}(T)$ for a given temperature, as illustrated in the upper right of Fig. 3(d). The $H_{c1}(T)$ values at different temperatures plotted in Fig. 3(d) are used to determine $H_{c1}$(0) of 6.9$\pm$0.1 mT through the empirical relation \cite{Tin} $H_{c1}(T)/H_{c1}(0)=1-(T/T_c)^2$. With these critical fields, $\lambda_L$ and the Ginzburg-Landau coherence length ($\xi_{GL}$) can be determined by using the equations: $H_{c2}(0)=\Phi_0/2\pi\xi_{GL}^2$ and $H_{c1}(0)=(\Phi_0/4\pi\lambda_L^2)\ln(\lambda_L/\xi_{GL})$ with the flux quantum $\Phi_{0}=2.0678\times10^{-15}$ Wb. We thus have $\lambda_L$=3325$\pm$36 {\AA}, $\xi_{GL}$=31.6$\pm$0.3 {\AA}, and the Ginzburg-Landau parameter $\kappa=\lambda_L/\xi_{GL}=105\pm2$. These results are compared with those determined by using various techniques such as the magnetization \cite{Hol,Iro,Bun,wong}, magnetoresistance \cite{Hou}, muon spin relaxation \cite{Uem,Uem1}, nuclear magnetic resonance \cite{Tyc,Sas1}, and optical reflectivity \cite{Deg}.

\begin{figure}[tbp]
\includegraphics[width=\columnwidth]{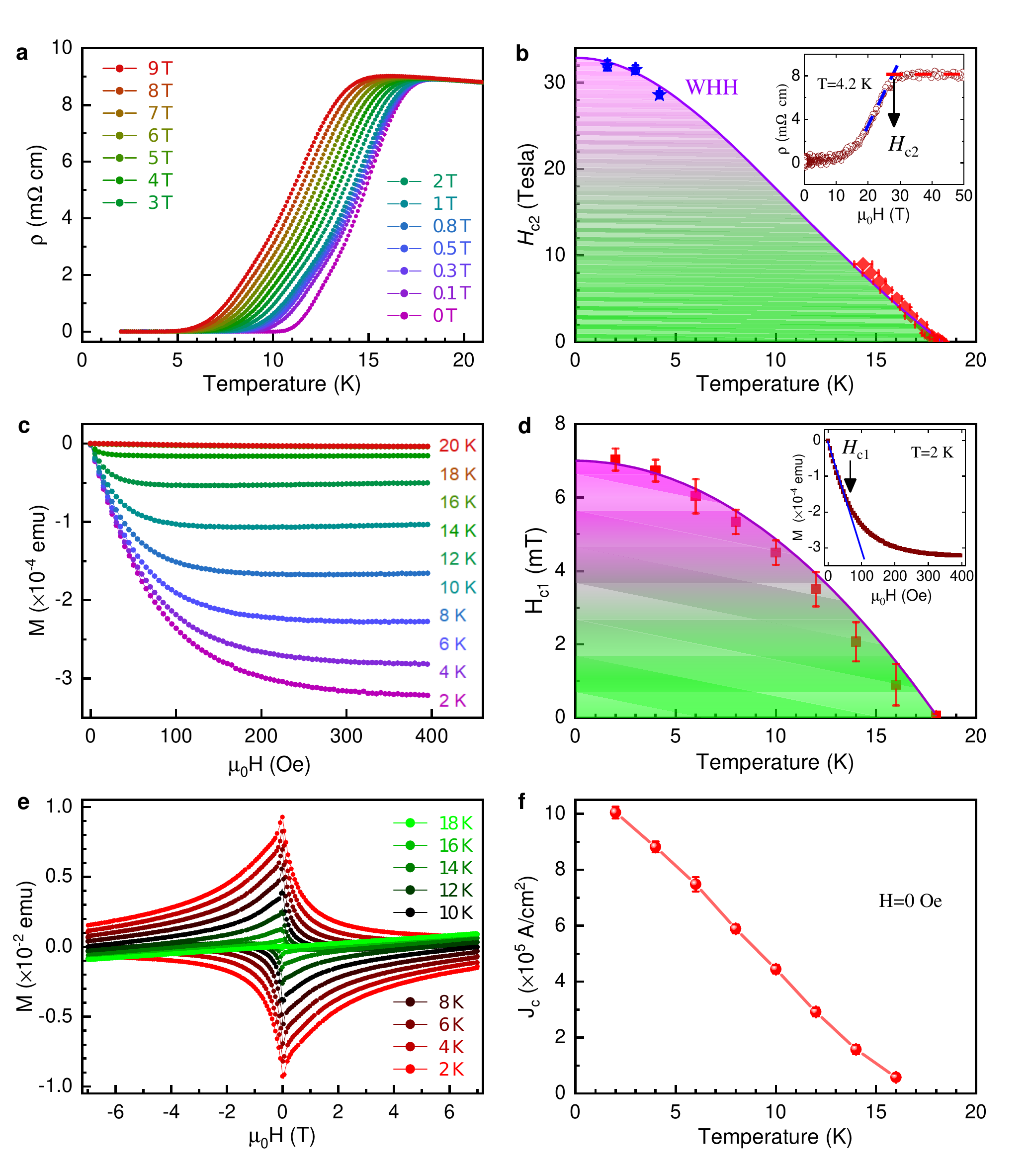}
\caption{Determination of superconducting parameters of K$_3$C$_{60}$. (a) Temperature-dependent resistivities at various magnetic fields up to 9 T. (b) Temperature dependence of the upper critical field $H_{c2}(T)$. The data points of the red diamonds and blue pentacles are obtained from the electrical transport measurements based on the fixed-field (a) and fixed-temperature scans, respectively. Error bars represent estimated uncertainties in determining $H_{c2}$. The solid line is the fitting to the Werthamer-Helfand-Hohenberg theory. Inset: Magnetic field dependent resistivity with scanning field up to 50 T at temperature of 4.2 K. $H_{c2}$ for a fixed temperature is determined from the intercept of the linear extrapolations from below and above the transition. (c) Magnetic-field dependence of the magnetization ($M$) at various temperatures. (d) Temperature dependence of the lower critical field $H_{c1}(T)$. The solid curve is the fitting of the measured data points to the empirical law $H_{c1}(T)/H_{c1}(0)$=1$-$($T/T_c$)$^2$. Inset: The magnetization as a function of magnetic field at 2 K. The derivation from the initial linear trend is used for the determination of $H_{c1}$. (e) Magnetic hysteresis ($\emph{M-H}$) loops at various temperatures with the applied field up to $\pm$7 T. (f) Temperature dependence of the critical current density ($J_c$). The $J_c$ value is determined from the $\emph{M-H}$ loop (e) based on the Bean's critical-state model.}  
\end{figure}

\subsection{Determination of critical current density}

Figure 3(e) displays the $dc$ magnetic hysteresis loops at several fixed temperatures below $T_c$. With increasing temperature, the hysteresis loops gradually shrink inward until converting to a straight line near $T_c$. The diamond-like loop and temperature-dependent shrinkage are typical features for a type-II superconductor. The critical current density ($J_c$) can be determined simply from these $dc$ magnetic loops based on Bean's critical state model \cite{Bea} by using the formula $J_c=A\times(M_+-M_-)/r$, where M$_+$ and M$_-$ are the magnetizations in the decreasing and increasing circles at a given field $H$, and $A$ and $r$ are the shape of sample \cite{Fie} and the sample radius, respectively. In our estimation,  the radius $r$ is assumed to be 1 $\mu$m according to the XRD results. The $J_c$($H$=0) variation with temperature is shown in Fig. 3(f). The obtained $J_c$ value for K$_3$C$_{60}$ in the present work differs largely with the early results obtained from the same magnetization measurements \cite{Hol,Bos,Zha,Iro,Bun,Lee,Bun1,Bun2,Bae}. The difference mainly comes from the uncertainties for the radius $r$, which has been confirmed by the functional relationship between $J_c$ and the particle size \cite{Bos}. Meanwhile, the granularity of the sample also affects the $J_c$ value \cite{Bun}. 

It should be emphasized that $J_c$ is important for evaluating technological applications of a superconductor. The material itself possesses a self-field component $J_c$(s) even without applying external field. For a type-II superconductor, $J_c$(s) follows a universal expression \cite{Tal} $J_c$(s)=$H_{c1}$/$\lambda_L$. It is clear that $J_c$(s) is independent of the material geometry. Substituting the obtained $H_{c1}$ and $\lambda_L$, we have $J_c$(s) of 1.7$\times$10$^6$ A/cm$^2$ for K$_3$C$_{60}$, slightly larger than $J_c$($H$=0) determined from the magnetization measurements [Figs. 3(e)-3(f)]. Therefore, the technological parameters of $T_{c}$, $H_{c2}$, and $J_{c}$ needed for K$_{3}$C$_{60}$ to function as a superconducting magnet material are well established. 

\begin{figure}[tbp]
\includegraphics[width=1.0\columnwidth]{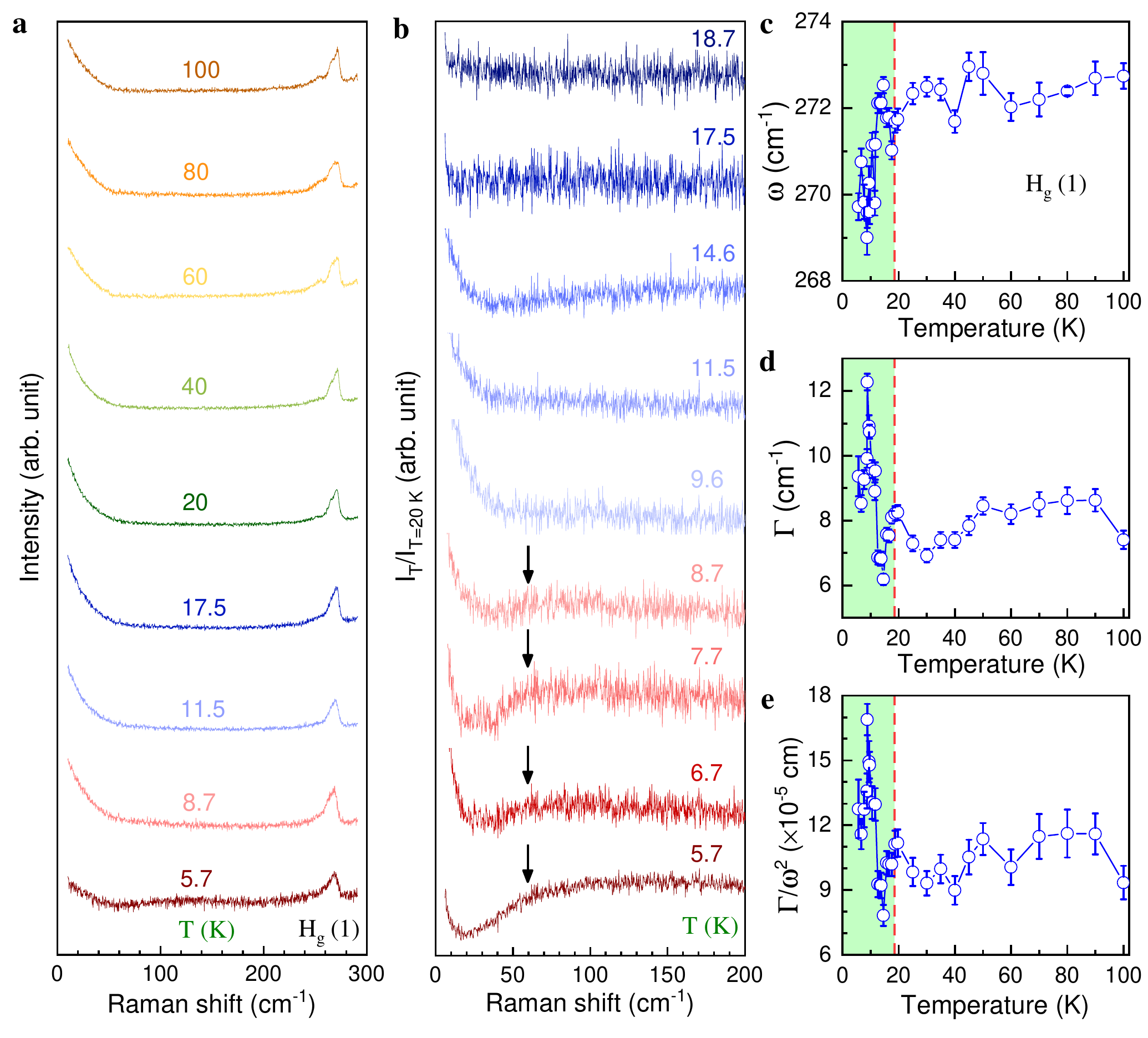}
\caption{Raman spectra in the low-frequency region of K$_3$C$_{60}$. (a) Raman spectra at several representative temperatures before and after the superconducting transition. The scattering peak at around 270 cm$^{-1}$ is assigned as the $H_g$ (1) mode. (b) Raman spectra normalized by the intensity at 20 K. The intensity shows a linear decrease below 60 cm$^{-1}$ (as marked by the arrows) when the temperature is lower than 1/2 $T_c$, and it becomes more obvious at lower temperatures. (c-e) Temperature dependence of the frequency ($\omega$), linewidth ($\Gamma$), and electron-phonon coupling parameter $\Gamma/\omega^2$ of the $H_g(1)$ mode, from top to bottom, respectively. Green shadow areas represent the superconducting state.}
\end{figure}

\subsection{Detection of superconducting gap and evaluation of electron-phonon coupling}
 
The superconducting energy gap ($\Delta$) of K$_3$C$_{60}$ is determined based on the Raman spectroscopy measurements. Figure 4(a) shows the representative Raman spectra in the low-frequency region at a wide temperature range from 5.7 to 100 K. In Fig. 4(b), the Raman spectra at low temperatures below $T_c$ were normalized by dividing the spectrum measured at temperature of 20 K. One can see clearly a linear decrease in the scattering intensity below 60 cm$^{-1}$ (as marked by arrows) when the temperature cools down to 1/2 $T_c$. This feature becomes more obvious at lower temperatures. Such a linear decrease in the intensity of the electronic scattering can be attributed to the renormalization of the density of state due to the superconducting transition \cite{Els}. The 2$\Delta$ value can be taken from the position where the frequency starts to deviate the linear behaviour. The reduced energy gap $2\Delta/k_BT_c$ of 4.7$\pm$0.3 is thus obtained with $k_B$ being the Boltzmann constant. This value is higher than 3.53 predicted within the BCS framework. The high $2\Delta/k_BT_c$ and large $H_{c2}(0)$ together indicate the strong coupling of the interactions in this superconductor. Similar conclusions have also been drawn from other techniques including point-contact tunneling spectroscopy \cite{Zhang,Ren} and nuclear magnetic resonance \cite{Sas}. Some other studies based on the optical reflectivity \cite{Deg,Deg1}, nuclear magnetic resonance \cite{Tyc,Sas}, and photoemission \cite{Hes} support weak electron-phonon coupling within the conventional BCS framework. 

For type-II spin-singlet superconductors, the orbital and Pauli paramagnetic effects are two distinct ways for pair-breaking with increasing external magnetic field. Using $2\Delta/k_BT_c$=4.7$\pm$0.3, the Pauli-limiting field $H_P$=45.4$\pm$2.2 T is then determined accurately from $H_P=\Delta/(\sqrt{2}\mu_B)$ \cite{Clo} with $\mu_B$ being the Bohr magneton. The obtained orbital component of the upper critical field $H_{c2}^{orb}$ of 22 T is much smaller than $H_P$, yielding the Maki parameter \cite{Gru} $\gamma=\sqrt{2}H_{c2}^{orb}/H_P\sim0.7$. Since $\gamma$ reflects the strength of the paramagnetic effect, this indicates that the orbital effect dominates the $H_{c2}$ behaviour for this superconductor.

\begin{figure}[tbp]
\includegraphics[width=1.0\columnwidth]{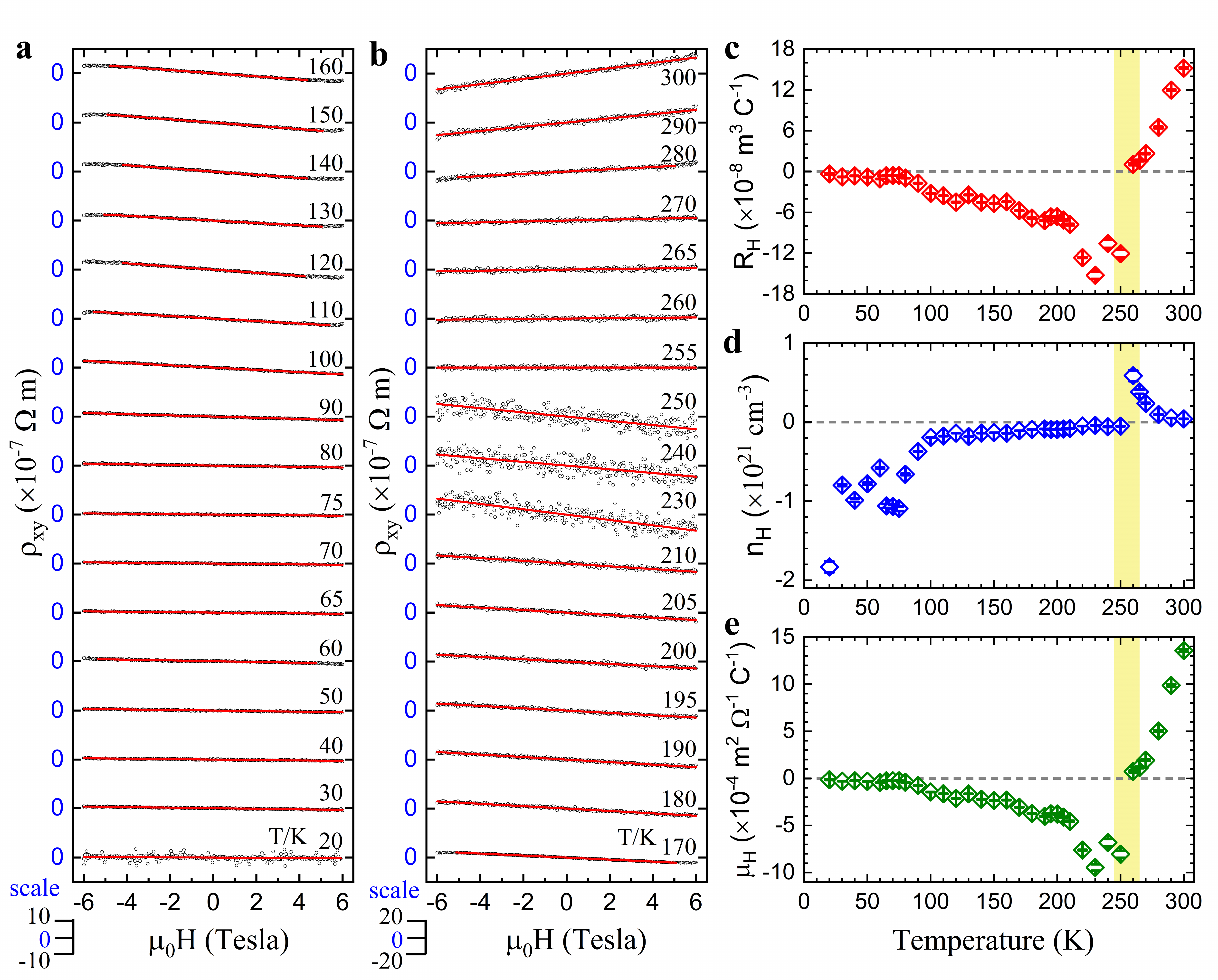}
\caption{Hall effect data of K$_3$C$_{60}$. Hall resistivity ($\rho_{xy}$) vs. the applied magnetic field along the two opposite directions up to 6 T in the temperature range from 20 to 160 K (a) and from 170 to 300 K (b). Numerical scale is given in the left bottom. The Hall coefficient ($R_H$) (c) carrier concentration ($n_H$) (d) and mobility ($\mu_H$) (e) as a function of temperature. The yellow regions in (c-e) denote the boundary for the occurrence of the orientational transition.}
\end{figure}

In addition, the electron-phonon contribution to the superconductivity can be evaluated from the observations of the Raman spectra at temperatures below $T_c$. The low-frequency $H_g$(1) intramolecular vibrational mode is chosen as an example to monitor the superconductivity-induced changes. The systematic change of this mode can be seen from Fig. 4(a). The temperature dependence of the frequency ($\omega$), linewidth ($\Gamma$), and electron-phonon coupling parameter $\Gamma/\omega^2$ of this $H_g$(1) mode is shown in Figs. 4(c)-4(e), from top to bottom, respectively. When entering the superconducting state (the green shaded zone), the $H_g$(1) mode exhibits the downshift (softening) in $\omega$, the increase (widening) in $\Gamma$, and the enhancement of $\Gamma/\omega^2$. These behaviours are due to the superconductivity-induced phonon self-energy effect. The early proposal \cite{prass} regarding the phonon evolution with temperature when this material enters the superconducting state with high-resolution from an inelastic scattering technique is finally realized here. The absence of phonon mode near 40 cm$^{-1}$ in the Raman spectra at temperatures below $T_{c}$ indicates that it is probably an acoustic branch and thus inactive to Raman scattering but visible in the neutron measurements \cite{prass}. These results provide the strong support for the important contribution of the electron-phonon coupling to the superconductivity in K$_3$C$_{60}$. The observed self-energy effect together with the large $2\Delta/k_BT_c$ obtained from the same Raman spectra reveals the strong electron-phonon coupling in K$_3$C$_{60}$, which needs to be included in dealing with the phonon effects \cite{varma,fczhang,schl,chen,han,Nom,dolg}.  

\begin{table}[tbp]
\caption{Summary of the superconducting parameters of K$_3$C$_{60}$.}
\begin{center}
	\begin{tabular}{cccccc}
		\hline
		\hline
		 $T_c$ (K) &&& && 18.5$\pm$0.5\\ 
	   $H_{c1}$(0) (mT) &&& & &6.9$\pm$0.1\\ 
	   $H_{c2}$(0) (T)	&&& && 33.0$\pm0.5$ \\
	   $\lambda_L$ ({\AA}) &&& & &3325$\pm$36 \\
	   $\xi_{GL}$ ({\AA}) &&& & &31.6$\pm$0.3 \\
	   $J_c$ ($\times$10$^4$ A/cm$^2$) &&& && 100$\pm$2\\
	  2$\Delta$/$k_BT_c$ && && &4.7$\pm$0.3 \\
	   $v_F$ ($\times$10$^{4}$ m/s) &&&& & 5.6$\pm$0.4\\
	   $E_F$ (meV) & &&& &55$\pm$13\\
	   $T_F$ (K) && &&& 641$\pm$147\\ 
		\hline
		\hline
	\end{tabular}
\end{center}
\end{table}

\subsection{Determination of Hall coefficient}

The magnetic-field-dependent Hall resistivities of K$_3$C$_{60}$ with applied magnetic fields up to $\pm$6 T at various temperatures from 20 to 300 K are shown in Figs. 5(a) and 5(b). The Hall resistivity $\rho_{xy}$ versus $H$ curves at all studied temperatures are essentially linear, ensuring the accurate determination of the Hall coefficient $R_{H}$ through the linear fitting to the data points [Fig. 5(c)]. The carrier concentration $n_{H}$ and mobility $\mu_{H}$ are thus obtained through $n_{H}$=$1/(eR_H)$ and $\mu_{H}$=$\sigma_{xx}R_{H}$ with $e$ and $\sigma_{xx}$ being the electron charge and the electrical conductivity, respectively [Figs. 5(d)-5(e)]. As can be seen, $R_{H}$ changes sign at the temperature range of 220-250 K. The change of the carrier character further indicates that the synthesized sample is in a half-filled state with both electron and hole conduction \cite{Pal,Erw}. Similar sign change has also been observed in K$_3$C$_{60}$ thin films \cite{Pal} but was absent in the study for K- and Rb-doped C$_{60}$ single crystals \cite{Lu}. Interestingly, around the temperature for the sign change, the orientational ordering transition has been reported for C$_{60}$ \cite{Hei} and K$_3$C$_{60}$ \cite{rswang,Bar}. The Hall effect clearly captures the effect of the orientational ordering on the electronic structure. Most importantly, the obtained evolution of $n_H$ with temperature close to $T_{c}$ not only gives the accurate carrier concentration for K$_3$C$_{60}$ to superconduct but also determines the electronic feature at low temperatures rather than the hole character at room temperature, though the two carrier behaviour was noticed in the band structure calculations \cite{Erw} and the electron doping is generally believed.  

\begin{figure}[tbp]
\includegraphics[width=1.0\columnwidth]{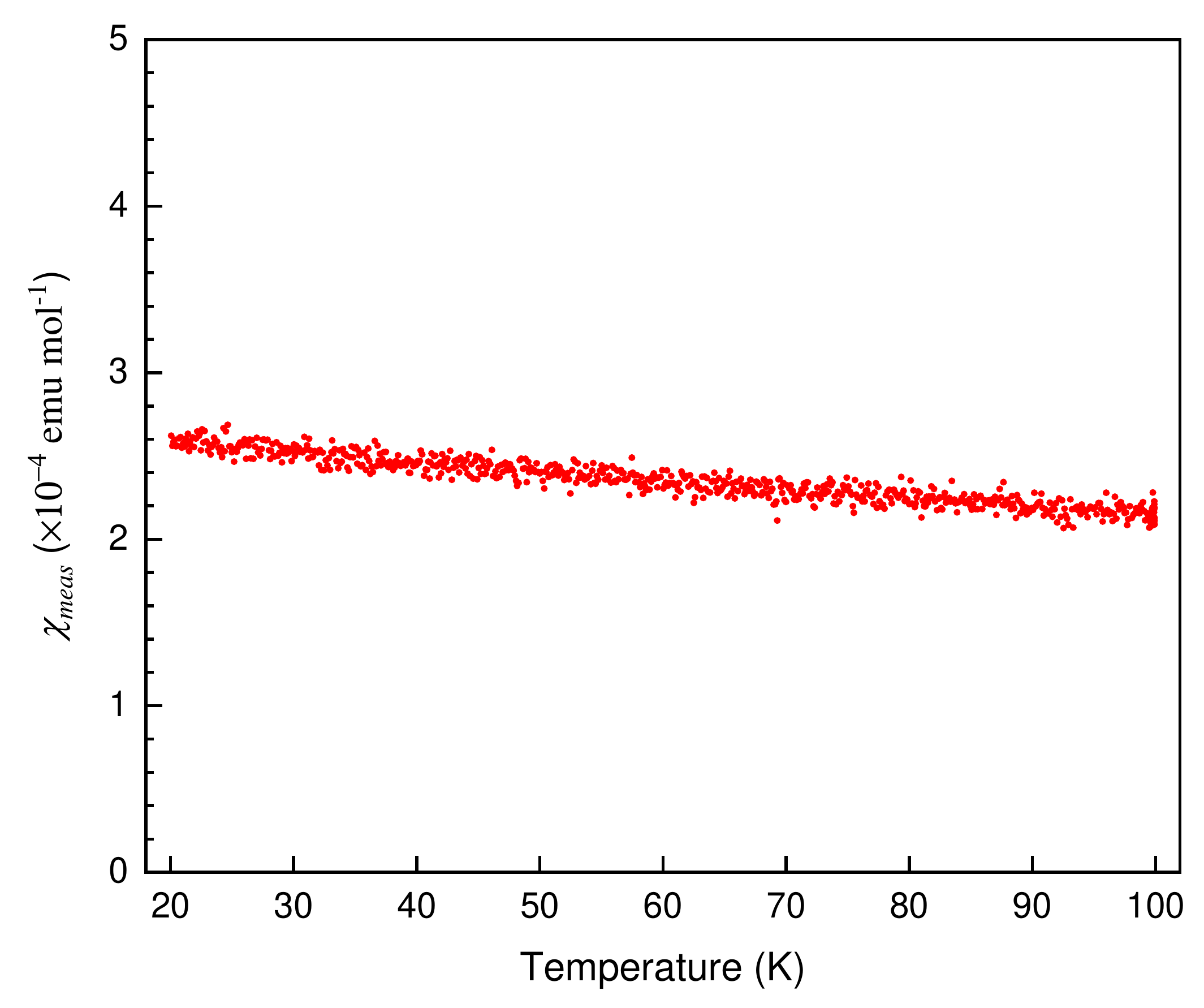}
\caption{Magnetic susceptibility $\chi_{meas}$ of K$_3$C$_{60}$ in the temperature range of 20-100 K.}
\end{figure}

\subsection{Full set of superconducting parameters}
           
Table I summarizes the obtained superconducting parameters for K$_3$C$_{60}$. These parameters obtained from the same sample could settle down the large contradictions from different groups. For instance, the reported $H_{c1}$(0) values in the literature are in the range from 1 to 13 mT with the difference in one order of magnitude \cite{Hol,Iro,Bun,Bun3}. Our $H_{c1}$(0) value locates in the middle of this range. In the case of $2\Delta/k_BT_c$, the early results are in the range of 3-3.6 revealed by far-infrared  the optical reflectivity \cite{Deg,Deg1}, nuclear magnetic resonance \cite{Tyc,Sas}, and photoemission \cite{Hes}, suggestive of the weak BCS coupling. Meanwhile, the strong coupling effect is supported by the large $2\Delta/k_BT_c$ values from 4.3 to 6 based on the scanning tunneling microscopy \cite{Zhang,Ren} and nuclear magnetic resonance \cite{Sas} measurements. The obtained value of 4.7$\pm$0.3 for K$_3$C$_{60}$ from the present work is in the middle range of these studies, which supports the strong coupling effect. 

Having the knowledge of $\xi_{GL}$ based on the well determined $H_{c2}$(0) as well as $\Delta$, one can derive the average Fermi velocity $v_F$ of (5.6$\pm$0.4)$\times$10$^{4}$ m/s by using the formula $v_F=\pi\Delta\xi_{GL}/\hbar$, in good agreement with the measurements and calculations \cite{wong,Erw}. The measured magnetic susceptibility $\chi_{meas}$ (Fig. 6) at 20 K gives the Pauli susceptibility $\chi_{spin}$=(7.5$\pm$0.6)$\times$10$^{-4}$ emu/mol after the correction of the core and Landau diamagnetic contributions. Taking the $n_H$ value at 20 K (Fig. 5) for the conduction electron density $n$, we obtain the effective mass $m^{\ast}$=(6.2$\pm$0.6)$m_0$ from the expression $\chi_{spin}$=$\mu_Bm^{\ast}(3\pi^2n)^{1/3}/\hbar^2\pi^2$, where $m_0$ and $\hbar$ are the mass of the free electron and the reduced Planck constant, respectively. The result for $m^{\ast}$ is very close to (6.4$\pm$1.5)$m_0$ in the early study \cite{wong}. Using the $v_F$ and $m^{\ast}$ value, we have $E_F$=55$\pm$13 meV and the Fermi temperature $T_F$=641$\pm$147 K for K$_{3}$C$_{60}$ based on the formula $E_F$=$m^{\ast}$$v_F^2$/2=$k_B$$T_F$. Note that all these important parameters are obtained on the same sample in terms of the generally accepted techniques. Table I is thus updated to include them. 

\section{Discussion and conclusions}
 
It was often expected to elucidate the driving force for superconductivity from the isotope effect. However, the existing isotope exponent $\alpha$ deviates the predicted value of 0.5 from the BCS theory and can be divided two groups depending on the $^{13}$C enrichment, one \cite{Che1} with $\alpha$$<$0.5 and the other \cite{Zak,Aub,Ric} with $\alpha$$>$1.0. The former suggests the strong coupling effect, consistent with our Raman data, and the latter favors the electron correlations \cite{chak2,ashc}. Therefore, the isotope effect suggests that the superconducting pairing in K$_{3}$C$_{60}$ originates from either the strong electron-phonon interaction or the electronic correlations. However, it is hard to pin down the exact mechanism due to the scatter of the data.    

\begin{figure}[tbp]
\includegraphics[width=1.0\columnwidth]{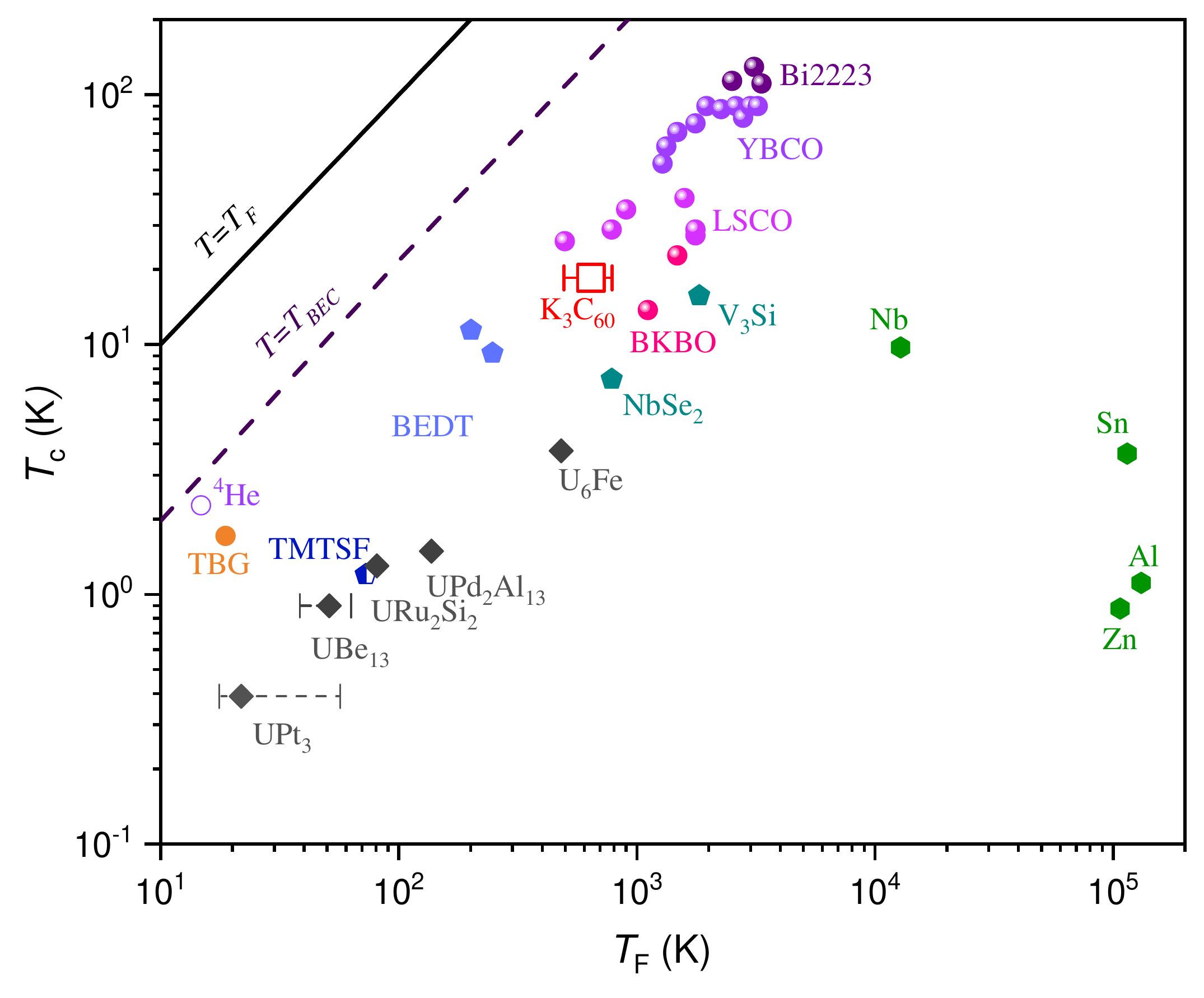}
\caption{Relationship between the critical temperature $T_c$ and the Fermi temperature $T_F$ of representative superconductors. The data derived from the current measurements for K$_3$C$_{60}$ (open square) are compared with those of various superconductors taken from the previous works \cite{ycao,Uem3}. Bi2223, YBCO, LSCO, and BKBO represent Bi$_2$Sr$_2$(Ca,Pb)$_2$Cu$_3$O$_{10+\delta}$, YBa$_2$Cu$_3$O$_{7-\delta}$, La$_{2-x}$Sr$_x$CuO$_4$, and Ba$_{1-x}$K$_x$BiO$_3$, respectively. TMTSF and BEDT are two organic charge-transfer salts. TBG denotes twisted bilayer graphene. The broken line denotes the Bose-Einstein condensation temperature $T_{BEC}$ for an ideal three-dimensional boson gas.}
\end{figure}

Now that a full set of parameters for K$_{3}$C$_{60}$ (Table I) has been obtained from the same sample at all the necessary measurement conditions and techniques, they should provide the reliable constraints on the examination of the existing theories and the future theory development for the mechanism of superconductivity. The determined $H_{c2}$(0) not only yields $\xi_{GL}$ and but also further gives $v_F$ with the help of $\Delta$. The magnetization and Hall effect data give $m^{\ast}$ and thus $T_F$ when combining $v_F$. These self-consistently obtained parameters along with $T_{c}$ at least can be used to see how far away or close the studied K$_{3}$C$_{60}$ locates to the well known superconductors. 

A recognized measure for judging the nature of correlated superconductors is through the change trend for $T_c$ and $T_F$. All the reported correlated superconductors discovered so far including cuprates, heavy fermion systems, organic materials \cite{Uem2,Uem3}, and even twisted graphene \cite{ycao} follow a linear relationship between $T_c$ and $T_F$. The common characters for these superconductors are the existence of some competing order(s) and the strong electron correlation effect. Thus, the electron correlations are believed to be the major player for superconductivity in these systems. When adding the obtained $T_c$ and $T_F$ values for K$_3$C$_{60}$ in the map (Fig. 7), we find that this superconductor nicely follows the same linear trend as the other correlated superconductors but being very close to Ba$_{1-x}$K$_x$BiO$_3$. The trend itself classifies K$_3$C$_{60}$ as an unconventional correlated superconductor. The neighbourhood for Ba$_{1-x}$K$_x$BiO$_3$ and K$_3$C$_{60}$ implies that their superconductivity may be governed by the similar factors. For Ba$_{1-x}$K$_x$BiO$_3$, till recently it was found \cite{Wen} that the electron correlations enhance the electron-phonon coupling and they together contribute to the high $T_c$ in this system. It is reasonable to believe both the electron correlations and electron-phonon coupling jointly account for superconductivity in K$_3$C$_{60}$. This idea finds the support from the theoretical calculations on alkali-doped fullerides mainly focusing on Cs$_{3}$C$_{60}$ \cite{Nom}, where the electron-phonon coupling was found to enhance the electron correlations and they together gave the correct $T_{c}$ evolution trend with the unit cell volume, though the calculated $T_{c}$ is generally lower about 10 K than the experimental value. Since the strong coupling effect in K$_3$C$_{60}$ has been firmly established from high $T_c$, high $H_{c2}$(0), large $2\Delta/k_BT_c$, and the Raman data, the $T_c$ and $T_F$ relation derived from other parameters further suggests unconventional correlated superconductivity and the joint contributions from the electron-phonon coupling and electron correlations \cite{Gun0,capo}. Recalling the unconventional isotope effect, the nice collaboration between the strong electron-phonon coupling and electron correlations is thus suggested to be responsible for the observed unconventional superconductivity in K$_3$C$_{60}$, both in a favorable way. This should be the base and constraints on the examinations of the existing theories and the future theory development for the mechanism of superconductivity for fullerides. 

\section{EXPERIMENTAL DETAILS}

\noindent\textbf{Sample synthesis.} High-quality K-doped C$_{60}$ sample was synthesized by using a modified version of a solution-phase reaction process detailed previously \cite{Yoo,rswang,zong}. High-purity potassium (99\%, Sinopharm Chemical Reagent) and fullerene powder (99.9\%, Acros Organics) were mixed with the nominal mole ratio of 3:1 (K:C$_{60}$). The mixtures, together with certain dose of ultra-dry tetrahydrofuran (THF) solvent, were loaded into a 15 ml borosilicate vial. The glass vial was then treated in an ultrasonic device at around 50 $^{\circ}$C for 5-10 minutes to accelerate the dissolution of alkali metals and the reaction between potassium and C$_{60}$ molecules. The colour of the solution turned reddish-brown after the ultrasound process. It should be noted that the temperature of the reaction can not exceed 63 $^{\circ}$C, which is the melting point of potassium, otherwise potassium would melt into small spheres that are difficult to react, thus breaking the desired stoichiometric ratio for superconducting phase. After that, the solution was oscillated in a vortex mixer (Vortex 3000, Wiggens) with 200 r/min for 10 hours to ensure the complete reaction and the homogeneity of sample. Upon filtering, we obtained a black preliminary product. All the above preparation steps were carried out in an argon-filled glovebox with both O$_{2}$ and H$_{2}$O levels less than 0.1 ppm. The as-prepared K$_{3}$C$_{60}$ was then put into quartz tubes and sealed under vacuum about 1$\times$10$^{-4}$ Pa. The fragile K$_{3}$C$_{60}$ sample was obtained after annealing at 250 $^{\circ}$C for 20 hours.

\noindent\textbf{Characterization of crystal structure.} The crystal structure was determined based on the X-ray diffraction spectrometer (Panalytical Emperean) by using Cu K$_{\alpha}$ radiation with wavelength of 1.5406 {\AA}. The polyimide film was used to cover the sample to avoid the oxidation of K$_{3}$C$_{60}$. The background signals of the film were carefully subtracted in the subsequent analysis. The space group and lattice parameters were determined by using $\emph{Jana 2020}$ program \cite{jana} based on the Le Bail method \cite{bail} to fit the diffraction patterns.

\noindent\textbf{Raman spectroscopy measurements.} The Raman specta were collected in an in-house system with Charge Coupled Device and Spectrometer from Princeton Instruments. The laser with the wavelength of 488 nm and power less than 2 mW was used in the measurements. Pristine and doped C$_{60}$ were sealed in capillary tubes when collecting the Raman spectra at room temperature [Fig. 1(c)]. For cryogenic Raman spectroscopy experiment, the doped sample was loaded into a sealed cooper holder with an optical quartz window. The sample was then put on the holder equipped with a heater in a cryogenic vacuum chamber to obtain the temperature-dependent Raman spectra (Fig. 4).

\noindent\textbf{Magnetization measurements.} The magnetization measurements were carried out by using a Magnetic Properties Measurement System (MPMS3, Quantum Design). The sample was placed into a nonmagnetic capsule and sealed by GE Varnish to protect K$_{3}$C$_{60}$ from air. The $dc$ magnetic susceptibility $\chi(T)$ curves were collected with the ZFC and FC runs at the field of 10 Oe at temperature ranging from 1.8 to 100 K. In the $ac$ magnetic susceptibility measurements, the used probe harmonic magnetic field and frequency are 5 Oe and 234 Hz, respectively. The $M(H)$ plots at various fixed temperatures were collected by two steps. Firstly, the step-size was set to 5 Oe in stable mode from 0 to 400 Oe. Secondly, the $H$ swept from +7 T to -7 T, and then went back to get the complete hysteresis loop.

\noindent\textbf{Pauli (spin) susceptibility $\chi_{spin}$.} The measured magnetic susceptibility $\chi_{meas}$ is determined by the sum of the paramagnetic ($\chi_P$) and diamagnetic ($\chi_D$) components. The $\chi_D$ component includes the core and/or conduction electron (Landau) contributions, where the former can be estimated based on the published Pascal constants, and the latter usually takes 1/3 for the paramagnetic term. The value of $\chi_D$ for K$_3$C$_{60}$ can be calculated by $\chi_D$(K$_3$C$_{60}$)=60$\chi_D$(C atom)+3$\chi_D$(K atom)+$\chi_{Landua}$(conduction electron), in which $\chi_D$(C atom)=-6.0$\times$10$^{-6}$ emu/mol and $\chi_D$(K atom)=-18.5$\times$10$^{-6}$ emu/mol. We thus obtain $\chi_D$=-(5.0$\pm$0.3)$\times$10$^{-4}$ emu/mol. By using $\chi_{meas}$=(2.5$\pm$0.3$)\times$10$^{-4}$ emu/mol at 20 K (Fig. 6), we have the total bulk paramagnetic susceptibility $\chi_P$=(7.5$\pm$0.6)$\times$10$^{-4}$ emu/mol, which can be taken as the Pauli paramagnetic susceptibility $\chi_{spin}$.

\noindent\textbf{Resistivity and Hall effect measurements.} Due to the high sensitivity of the sample to air, a nonmagnetic Ni-Cr-Al alloy cell equipped with four air-tight copper leads was developed to measure the electrical and Hall resistivities when keeping good contact between sample and four electrodes and avoiding oxidation. The resistivity and Hall coefficient were determined in terms of the van der Pauw method \cite{pauw}. The sample thickness for the electrical transport measurements is 0.2 mm. The low-field resistivity and Hall effect measurements were performed on Physical Property Measurement System (Quantum Design, PPMS). The resistivities at paulsed magnetic fields up to 50 T with the help of a typical four-contact method were measured at the National High Magnetic Field Center, Wuhan, China.

\section*{Acknowledgements}

This work was supported from the Shenzhen Science and Technology Program (Grant No. KQTD20200820113045081), the Basic Research Program of Shenzhen (Grant No. JCYJ20200109112810241), and the National Key R$\&$D Program of China (Grant No. 2018YFA0305900).

\end{document}